\documentstyle[aps,epsfig,prl,twocolumn]{revtex}

\begin{document}

\vspace{0.0cm}
\draft

\title{Comment on ``Black holes constrain varying constants''}
\author{V.V. Flambaum \thanks{email address:
flambaum@newt.phys.unsw.edu.au}}

\address{Institute for Nuclear Theory,  University of Washington,
Seattle WA, 98195-1550, and School of Physics, University of New South Wales,
Sydney 2052, Australia}

\maketitle
\vspace*{-1.0truecm}

\begin{abstract}
A recent paper \cite{DDL} claims that the increase of the proton electric
charge $e$ leads to a (forbidden) decrease of black hole
entropy, therefore, possible evidence for variation of
$\alpha=e^2/\hbar c$ \cite{alpha} must be interpreted as
a decrease of the speed of light.
We argue that purely theoretical consideration of black holes possibly
cannot give any model-independent limitations on variation of
fundamental constants.

\end{abstract}



The entropy $S$ is, in fact, a dimensionless function.
Indeed, we can measure temperature in more natural energy units
and take the Boltzmann constant $k=1$ or just consider $S/k$.
It is convenient to present the dimensionless entropy of a charged
black hole in terms of dimensionless parameters:
\begin{equation}
\label{S}
S=\pi [\mu + \sqrt{\mu^2 - Z^2 \alpha}]^2
\end{equation}
Here $\mu=M/M_P$, $M$ is the black hole mass,$M_P=(\hbar c/G)^{1/2}$ is the
Plank mass, $Z$ is the black hole charge in units of the proton charge.
This expression does not contain explicitly any parameters which have
dimension \cite{note} (speed of light, proton electric charge, etc.).
Therefore, in this case we may only discuss variation of two
dimensionless parameters: mass of the black hole in units of the
Plank mass ($\mu$) and $\alpha$ \cite{rotation}.
Similar arguments were presented in a recent comment \cite{duff}.

However, the question remains: can we extract any limitations on variation
of $\mu(t)$ and $\alpha(t)$?
We should recall that the event horizon area $A$ of the black hole
is quantized \cite{bekenstein}. Because of the relation between
the entropy $S$ and the horizon area $A$ one also can obtain
the entropy quantization:
\begin{equation}
\label{SA}
S=\frac{c^3}{4 G\hbar} A = \pi \gamma \cdot n
\end{equation}
where $\gamma$ is a numerical constant, $n$=1,2,... is an integer
(in fact, in Ref. \cite{bekenstein} a neutral black hole
has been discussed; for the validity of eq. (\ref{SA}) we possibly need
an extra condition $n >>1$).
This quantization condition gives us $\mu$ as a function of $\alpha$
and the number of quanta $n$:
\begin{equation}
\label{mualpha}
\mu=\frac{M}{M_P}=\frac{\gamma \cdot n +Z^2\alpha}{2 \sqrt{\gamma \cdot n}}
\end{equation}
One may compare this result with that for the hydrogen atom
where we have energy levels in atomic units $me^4/\hbar^2$ as a function
of the principal quantum number $n$. These levels also depend on
$\alpha$ if we include relativistic corrections. Actually, such a dependence
of transition frequencies on $\alpha$ in many-electron atoms
(this dependence was calculated in \cite{dzuba}) was used in \cite{alpha}
to measure the $\alpha$ variation. Therefore, the time dependence
of $\alpha$, possibly, does not lead to any specific problems for the
black holes, it just gives us the dependence $\mu(t)$.
The entropy in this case does not change.

We may even ignore this quantization condition and consider
a heavy ``classical'' black hole. One can suggest that the dependence
$\alpha (t)$ will lead to a zero or negative expression
under the square root sign in eq. (\ref{S}) which also appears
in the formula for the horizon area (see a similar suggestion
for variation of the charge $e$ in Ref. \cite{DDL}). Here again an atomic
analogy may be useful. The energy of the 1s level in the Dirac
equation in the Coulomb field is given by the following expression:
\begin{equation}
\label{E}
E= m c^2 \sqrt{1- Z^2 \alpha^2}
\end{equation}
Here the energy includes the rest mass $m c^2$.
If we take $Z=137\approx 1/\alpha$ and begin to increase $\alpha$
the expression under the square root becomes negative and the stationary state
1s disappears (for the nucleus of any finite size it does not happen at
energy $E=0$ but the problem appears again for $E= -mc^2$). This only
means that the system becomes non-stationary. The strong Coulomb field
creates an electron-positron pair, the positron goes to infinity
and the nuclear charge reduces to an allowed value ($Z$ to $Z-1$).
The same may happen to the black hole: after some radiation it
becomes ``normal''. Therefore, the possibility of a negative expression
under the square root sign in eq. (\ref{S}) or (\ref{E}) does not mean
that certain variations of $\alpha$ are forbidden.

This consideration provokes another question. Can Nature use variation
of fundamental constants to eventually eliminate black holes?
For an external observer the free fall of a body to the horizon
requires an inifinite time. If by that time the horizon disappears
we may arrive at some interesting problems. However, the job of elimination
of black holes, possibly, can be done more efficiently
by Hawking radiation.

Of course, this consideration is too simplistic and does not
involve any particular model of the fundamental constant variation.
However, it is clear that one can not obtain any model independent
limits  for the
separate variation of $e$ and $c$
from the consideration of  the black hole entropy
since  eq. (\ref{S}) for this entropy does not contain  $e$ and $c$.
Moreover, it does not seem to be easier to derive  reliable
limits on variation of the fundamental constants by considering
such complicated objects as black holes instead of usual atoms, nuclei
and stars.

I am grateful to Michael Kuchiev for stimulating discussions.


\begin{thebibliography}{99}
\bibitem{note}
Note that one can only measure cosmological (space-time) variation of the
dimensionless fundamental constants. Consider, for example, the
speed of light $c$. If we want to measure it in m/s we need
to dig out the dependence of the standard meter and the standard second
on the (variable) fundamental constants. After that we may
try to measure the time dependence of the dimensionless ratio\\
$\frac{c(t)}{meter(t)/second(t)}$.\\
However, it would be much simpler to measure $c$ in units $e^2/\hbar$.
This is equivalent to the measurement of the variation of $\alpha$.
\bibitem{rotation}
We may also consider a rotating black hole where\\
$S=\pi ([\mu + \sqrt{\mu^2 - Z^2 \alpha - J(J+1)/\mu^2}]^2 +J(J+1)/\mu^2)$\\
Here $J$ is the dimensionless (quantum) angular momentum.
\end{thebibliography}
\end{document}